\documentclass[pra,twocolumn,titlepage,nofootinbib,amsmath,showpacs]{revtex4}%
\usepackage{amsmath}
\usepackage{amsfonts}
\usepackage{amssymb}
\usepackage{graphicx}%
\begin{document}
\title{Constraints on a long-range spin-dependent interaction
from precision atomic physics}
\author{S.~G.~Karshenboim}
\email{savely.karshenboim@mpq.mpg.de} \affiliation{D.~I. Mendeleev
Institute for Metrology, St.Petersburg, 190005, Russia
\\ {\rm and}
Max-Planck-Institut f\"ur Quantenoptik, Garching, 85748, Germany}

\begin{abstract}
We present a phenomenological constraint on a pseudovector light
boson beyond the Standard Model, which can induce a long-range
spin-dependent interaction $\alpha^{\prime\prime}({\bf s}_1\cdot{\bf
s}_2)\,{\rm e}^{-\lambda r}/r$. In the range of masses from $4\;{\rm
keV}/c^2$ to those related to macroscopic distances (of
$\lambda^{-1}\sim1\;$cm) the spin-dependent coupling constant
$\alpha^{\prime\prime}$ of the electron-muon interaction is
constrained at the level below a part in $10^{15}$. The constraint
is weakened while extending to higher masses. The strongest
constraint is related to the lepton-lepton interaction. Constraints
on spin-dependent interactions of some other particles are also
discussed.  The results are obtained from data on the HFS interval
of the ground state in muonium and a few other light hydrogen-like
atoms.
\end{abstract}
\maketitle

\section{Introduction}

Muonium is a bound system, consisting of an electron and an
antimuon. Interaction between the constituents is dominated by the
Coulomb interaction. The major contribution to the
hyperfine splitting (HFS) also comes from one-photon exchange, but
from its magnetic part. The most accurate experimental values of the
related transitions are
\begin{eqnarray}
\nu (1s-2s)&=&2455\,528\,941.0(9.8)\;{\rm
MHz}\;,~~~[1]\;,\nonumber\\
\nu (1s, {\rm hfs})&=&4463\,302.776(51)\;{\rm
kHz}\;,~~~[2]\;.\label{exp}
\end{eqnarray}
Comparison of theory and experiments provides one of the most
accurate tests of quantum electrodynamics (QED) calculations (see,
\cite{EGS,my_rep} for details).

Since it is possible to control relativistic, recoil and QED
corrections, one can apply these data to check whether a leading
non-relativistic term obtained from the one-photon exchange is
consistent with the gross experimental picture and to constrain
possible deviations from the standard description.

The leading terms are
\begin{eqnarray}
\nu (1s-2s)&\simeq&\frac{3}{4}\,R_\infty\;,\nonumber\\
\nu (1s, {\rm hfs})&\simeq&E_{\rm F}=\frac{16\alpha}{3\pi}\,
\mu_{\rm B}\,\mu_\mu \,R_\infty\, m_e^2\;,\label{ef}
\end{eqnarray}
where we apply relativistic units in which $\hbar=c=1$,
$e^2/(4\pi)=\alpha$ is the fine structure constant, $m_e$ is the
electron mass and $R_\infty$ is the Rydberg constant, $\mu_{\rm
B}=e/(2m_e)$ is the Bohr magneton and $\mu_\mu$ is the muon magnetic
moment. Similar equations hold for related transitions in
other light hydrogen-like atoms.

To calculate the dominant contributions, one has at first to
determine the values of related fundamental constants, such as the fine
structure constant $\alpha$, the Rydberg constant $R_\infty$, the
muon magnetic moment $\mu_\mu$. The latter come from
a bunch of experiments involved in the data analysis
\cite{codata2006} and if a certain `new-physics' effect would affect
those experiments in a different way, the outcome could be
inconsistent.

In this paper we study a possible inconsistency in the interpretation of
different experiments caused by an exchange by a pseudovector boson
with a light mass and an ultraweak coupling.

Such a boson should affect the HFS interval, which has been measured
with high accuracy in a number of light two-body atoms.

The strongest constraint we derive here is from muonium physics,
on which we focus our attention and consider related data in
details. Weaker constraints from experiments on other atoms are
also considered.

The constraint on the exchange by an intermediate boson comes from
its one-particle-exchange (OPE) contribution. Meantime, even in the
OPE approximation there are other intermediate particles which
contribute to the effective muon-electron interaction and shift the
energy levels.

Some of such small contributions have been already included into
consideration of the muonium HFS as corrections due to weak and
strong interactions. At the tree level a $Z$ boson exchange should
be included (see, e.g., \cite{weak}). When taking into account
various perturbative effects and corrections one has to consider
exchange by $\rho$-meson, pion, $a_1$-meson etc. Those hadronic OPE
contributions are a result of certain reduction of more complicated
graphs to tree level. In particular, the $\rho$ exchange is taken
into account when considering the hadronic vacuum polarization
contribution \cite{rho}, while $\pi^0$ and $a_1$ exchange is a part
of consideration of the hadronic light-by-light-scattering
contribution \cite{lbl}.

In contrast to the situation with the anomalous magnetic moment of a
muon (see, e.g., \cite{muon}), the mentioned corrections ($Z$,
$\rho$, $\pi^0$, $a_1$) to the one-photon exchange are very small
and are of marginal importance for the present level of experimental
and theoretical accuracy in spectroscopy of muonium and other simple
atoms. Their smallness comes from the large masses of the
intermediate particles (all are much heavier than electron),
while in the hadronic case each induced vertex is additionally
suppressed, because it involves higher-order electrodynamics effects
and thus involves extra factors of $\alpha$.

Various unification theories (see, e.g. \cite{holdom,pospelov}) may
involve a particle lighter than an electron and here we constrain
its coupling constant.

A constraint on a possible new spin-independent interaction from
precision physics of simple atoms has been already considered
\cite{prl,previous} and here we discuss a spin-dependent
interaction, which is somewhat similar to those due to exchange by
$Z$, $\rho$, $\pi^0$, $a_1$. The characteristic momentum at the
ground state is about $4\;$keV and we consider an intermediate
particle lighter than that.

Two of mentioned interactions, namely due to $Z$ and $a_1$ exchange,
are of our particular interest, because they produce a spin-spin
coupling directly. By introducing such a spin-spin interaction, the
Coulomb exchange should be corrected
\begin{equation}\label{ssl}
-\frac{\alpha}{r} \to -\frac{\alpha+\alpha^{\prime\prime} \bigl({\bf
s}_e\cdot{\bf s}_{\overline{\mu}}\bigr){\rm e}^{-\lambda r}}{r}\;.
\end{equation}

Such a correction is specific for a pseudovector particles. For
instance, vector (e.g. a photon) or pseudoscalar (e.g. an axion)
particles produces a spin-spin interaction only as a relativistic
effect. In particular, in the case of spin 1/2 as long as the large
components of the Dirac wave function are considered the
spin-involving effects do not appear. The vector-particle exchange
is a kind of static electric interaction and the axion-induced
interaction is vanishing. Once we include the small components the
vector-particle exchange involve the magnetic forces and the axion
exchange becomes observable. Including of small components of both
interacting particles produces a suppression factor of
$1/(m_1m_2r^2)\ll1$, which dramatically weaken the spin-dependent
constraints on pseudoscalar particles (while for vector particles
more strong constraints originates indeed from spin-independent
interaction (see, e.g., \cite{previous}).

Studying the spin-spin interaction we can simplify (\ref{ssl}) under
certain conditions. In particular, if the intermediate particle is
massless (or, which is the same, lighter than $4\;$keV), the
resulting interaction at atomic scale should be
\begin{equation}\label{ss}
-\frac{\alpha}{r} \to -\frac{\alpha+\alpha^{\prime\prime} \bigl({\bf
s}_e\cdot{\bf s}_{\overline{\mu}}\bigr)}{r}\;.
\end{equation}

In principle, such a mechanism, involving a new particle, can
produce some ultraweak spin-dependent long-range interaction.

Here we consider an interaction, which is similar to $Z$ and $a_1$
exchange, but with different strength and mass. A possible range of
$\alpha^{\prime\prime}$ for the mass of the intermediate particle
(i.e., of the radius of the interaction) below $4\;$keV is under our
investigation \cite{prl}.

Such an intermediate particle is coupled to charged particles and
is rather expected to be unstable and to decay into photons.
However, as long as its width is much smaller than its mass, we can
consider the particle as stable while calculating the related
corrections to the energy levels (cf. with calculations of the $Z$
\cite{weak} and $a_1$ \cite{lbl} exchange for the muonium HFS).

\section{Method}

The HFS interval in light hydrogen-like atoms can be expressed in
terms of the so-called Fermi energy $E_{\rm F}$ and a correcting
factor due to reduced-mass, relativistic, recoil and QED effects.
Taking the latter into account \cite{EGS,my_rep} one can interpret
any measurement of the actual HFS interval as a measurement of
$E_{\rm F}$. The Fermi energy (see, e.g. (\ref{ef} for the muonium
Fermi energy) is in its turn proportional to a product of the muon
and electron magnetic moments.

The ground state HFS has been studied with a high accuracy in six
two-body atoms, which are muonium \cite{mu1shfs}, positronium
\cite{ps}, hydrogen \cite{exph}, deuterium \cite{expd}, tritium
\cite{expt} and helium-3 ion \cite{exphe}. The strength of the
constraint on a light pseudovector meson depends not only on
accuracy of the experimental determination of the HFS interval, but also
on accuracy of the theoretical calculation of this quantity. We briefly
overview the related theoretical problems in Sect.~\ref{s:other}.
Below, we focus our attention on muonium, study of which delivers
us the strongest constraint on $\alpha^{\prime\prime}$.


At present, a way of an $E_{\rm F}$ calculation is the following:
one takes an experimental value of $\mu_\mu$ and uses it in the
calculation. The value is obtained in macroscopic measurements, say
at $r>1\;$cm. If the interaction we are to constrain is related to
the case
\begin{equation}
1\;{\rm cm}^{-1}\ll\lambda\ll4\;{\rm keV} \;,
\end{equation}
then a certain mismatch, proportional to $\alpha^{\prime\prime}$ should
appear because the spin-spin term in (\ref{ss}) also contributes to
the HFS interval. The HFS interval is shifted by
\begin{equation}\label{hfscorr}
\delta E_{\rm
hfs}=-\frac{Z^2(\alpha+\alpha^{\prime\prime})^2m_r}{2}+\frac{Z^2\alpha^2m_r}{2}\;.
\end{equation}
This correction is universal for atoms with the nuclear spin 1/2.
(For the nuclear spin 1, e.g., in deuterium, a factor of 3/2 should
be introduced.) Here, $m_r$ is the reduced mass, which for all atoms
under study but positronium, is equal to the electron mass $m_e$
with a sufficient accuracy. In positronium, indeed, $m_r=m_e/2$.

The correction can be rewritten in terms of an effective correction
to magnetic moment in such a way that the Fermi energy with a
`corrected' magnetic moment includes a correction (\ref{hfscorr}).

In the case of muonium it is of the form
\begin{equation}
\delta E_{\rm hfs}= \frac{16\alpha}{3\pi} \mu_{\rm
B}\mu_\mu^{\prime} R_\infty  m_e^2\;,
\end{equation}
where $\mu_\mu^{\prime}$ is defined as
\begin{eqnarray}\label{mucor}
\mu_\mu^{\prime}
&=&-\mu_\mu\times\frac{2\alpha^{\prime\prime}}{\alpha}\,\frac{R_\infty}{E_{\rm F}}\nonumber\\
&=&-2.0\times 10^8\;\mu_\mu\;\alpha^{\prime\prime}\;.
\end{eqnarray}

That is the value of $\mu_\mu^{\prime}$ that should appear as a
mismatch in a determination of the muon magnetic moment, found from
a comparison of Eqs.~(\ref{ef}) and (\ref{exp}) and taking into
account all necessary reduced-mass, relativistic, recoil and QED
corrections \cite{my_rep,EGS,codata2006}, and a value, obtained by a
`direct' macroscopic measurement. Determination of magnetic moments
of various particles (muon, proton) and light nuclei (deuteron,
triton, helion (the nucleus of the helium-3)) is reviewed in detail
in \cite{codata2006} (see Sect.~VI there).

\section{Determination of the muon magnetic moment}

Let us consider a determination of the muon magnetic moment. The most
accurate value (in units of the proton magnetic moment) is
\cite{codata2006}
\begin{equation}\label{mp1}
\frac{\mu_\mu}{\mu_p}=3.183\,345 \,137(85)\;.
\end{equation}
The result is obtained after evaluation of all the world data. The
dominant contribution comes from a comparison of (\ref{ef}) and
(\ref{exp}) with all appropriate corrections taken, while the other
measurements are statistically negligible. That is not a value
obtained by any `direct' means.

If the correction (\ref{mucor}) is present, we should interpret this
result as
\begin{equation}\label{mp2}
\frac{\mu_\mu+\mu_\mu^{\prime}}{\mu_p}=3.183\,345 \,137(85)\;.
\end{equation}

This value should be compared with a `direct' measurement
\cite{mu1shfs}
\begin{equation}\label{mp3}
\frac{\mu_\mu}{\mu_p}=3.183\,345 \,24(37)\;.
\end{equation}
The latter is derived from a study of Breit-Rabi magnetic sublevels of
the ground state in the magnetic field. So, it is determined
from a macroscopic experiment.

The discussion on $\mu_\mu$ above involves  also $\mu_p$ (see Eqs.
(\ref{mp2} and (\ref{mp3})) as a unit. It appears in $E_F$, where
the actual dimensionless factor reads $\mu_\mu m_e/e =
(\mu_\mu/\mu_p)(\mu_p/\mu_{\rm B})(\mu_{\rm B} m_e/e)$. Note that
$\mu_{\rm B} m_e/e=1/2$, while the factor $\mu_p/\mu_{\rm B}$ is
determined from macroscopic experiments. Different scales are
related only to two determinations of $\mu_\mu$ and do not touch
any other involved quantities. The value of $\mu_p$ is customarily
involved in a presentation of the results but does not play any real
role in the issue under consideration.

For the references on measurements of these and similar quantities,
useful to examine HFS intervals in other light atoms, let us mention
that the magnetic moments of proton \cite{muep} and deuteron
\cite{mued} in units of the electron magnetic moment are determined
from a study of similar level structure as in muonium
\cite{mu1shfs}, while the  magnetic moments of triton
\cite{magtritium,nmr} and helion \cite{maghelium} are obtained from
NMR spectroscopy  (see also \cite{nmr} for an NMR determination of
$\mu_p/\mu_d$). To convert a result obtained in terms of $\mu_e$
into results in terms of $\mu_{\rm B}$, one has to apply a value of
$g_e$ measured in a macroscopic experiment \cite{aexp} as well.

\section{Constraining a long-range spin-spin interaction from muonium HFS}

The constraint on the electron-antimuon spin-dependent coupling
constant, resulting from comparison of (\ref{mp2}) and (\ref{mp3}),
reads
\begin{equation}\label{const}
\alpha^{\prime\prime}= \bigl(1.6\pm 6.0\bigr)\times 10^{-16}\;,
\end{equation}
which is the major result of the paper.

As already mentioned, the `direct' measurement deals with behavior of
hyperfine energy levels in macroscopic magnetic field
\cite{mu1shfs}. A value of $\mu_e/\mu_p$ has also been used as an
input datum that was obtained from measurements of splitting of
Breit-Rabi HFS sublevels in magnetic field \cite{muep}
(for an adequate theory see \cite{my_rep,codata2006,muepth}) at the
macsoscopic distance scale.

To be conservative, we estimate the distances from the field source
at the mentioned  macroscopic experiments as  larger than $1\;$cm.
In principle, one could consider a comparison of the atomic scale
with a somewhat larger distance. However, in this case it is
necessary to completely reanalyze both quoted experiments for their
magnetic effects, including the source of the field and the
shielding applied. In any case, from the point of view of particle
physics that is rather a higher-energy end for the scale which is of
interest. That range is related to the mass of an intermediate boson
of roughly $4\;$keV.

\section{Constraints from the ground state HFS interval in
other light H-like atoms\label{s:other}}

The constraint (\ref{const}) is related to a four-fermion
$e\overline{e}-\mu\overline{\mu}\;$ interaction. Some other
interactions can be also constrained from atomic physics. A similar
constraint can be also set for an $e\overline{e}-e\overline{e}\;$
interaction, but it is weaker by a few orders of magnitude. That is
because positronium HFS is much larger (i.e., the HFS is a bigger
portion of the Rydberg energy) and because of lower experimental
accuracy in the determination of the HFS interval \cite{ps}. The
constraint is at the level of a few parts in $10^{12}$. Since theory
and experiment disagree at the level of about $2.5\;\sigma$ (see,
e.g., \cite{my_rep}), perhaps, we have to estimate the experimental
and theoretical uncertainty somewhat more conservatively than in the
original publications. In any case, in such a specific area as a
constraint on `new physics', the aim is rather a conservative
limitation than a `detection', and some two- or three-sigma effects
are observed from time to time.

In the case of an $e\overline{e}-p\overline{p}\;$ interaction,
accuracy should be also somewhat lower than (\ref{const}) because of
relatively low theoretical accuracy, the uncertainty of which is due to
the proton structure effects (see, e.g., \cite{EGS,my_rep}). That is
compensated in part by a smaller value of the HFS interval because
of smaller value of the nuclear magnetic moment (cf. (\ref{mucor})).

A constraint for a compound particle can be derived, e.g., from the
deuteron HFS. The theoretical accuracy here is even worse than for
hydrogen  (see, e.g., \cite{my_rep}), but the enhancement because of
a small HFS interval is larger.

The exact value of constraints for $\alpha^{\prime\prime}$ for an
$e\overline{e}-p\overline{p}\;$ and $e\overline{e}-d\overline{d}\;$
interaction should come from an estimation of the
nuclear-structure-uncertainty. It seems, however, that a situation
with that for the HFS intervals in hydrogen and deuterium is
somewhat uncertain and we prefer to give a rough estimation. We
expect a constraint on the related $\alpha^{\prime\prime}$ values at
the level of a few parts in $10^{15}$. Utilization of data from other
light atoms is also possible.

A study of the HFS structure of other light atoms, such as tritium and
helium-3 ion, can provide similar constraints after the contribution
and uncertainty of their nuclear-structure effects are properly
estimated.

The results are summarized in Table~\ref{T:const1s}. We emphasize
that the constraints for non-leptonic atoms are rough estimations
and the accuracy of understanding of the nuclear structure effects
requires clarification.

\begin{table}[phtb]
\begin{tabular}{cr}
\hline
~~~~~Atom~~~~~ &  $\alpha^{{\prime\prime}}$~~~~~~~~~~  \\
 \hline
Mu &  $\bigl(1.6\pm 6.0\bigr)\times 10^{-16}$\\
Ps &  $(5.8\pm2.1)\times10^{-12}$\\
H &  $\pm1.6\times10^{-15}$\\
D &  $\pm8\times10^{-15}$\\
T &  $\pm7\times10^{-14}$\\
$^3{\rm He}^+$ &  $\pm5\times10^{-13}$\\
\hline
\end{tabular}
\caption{The constraint from the $1s$ HFS intervals on a coupling
constant $\alpha^{\prime\prime}$ for a pseudovector boson with mass
$\lambda\ll \alpha m_e \simeq 3.5\;$keV (which is related to the
Yukawa radius substantially above $a_0$).\label{T:const1s}}
\end{table}

All constraints, but the one from positronium, are based on a
comparison of a certain HFS interval and the related nuclear
magnetic moment, determined at macroscopic distances
\cite{mu1shfs,muep,mued,nmr,magtritium,maghelium,codata2006}. For
a specific case of positronium, the leading term for the HFS
interval
\begin{equation}
E_F({\rm Ps})=\frac76\,\alpha^2\,R_\infty\;.
\end{equation}
is calculated only from the  knowledge of electron charge and mass,
which enter in combinations, determination of which is insensitive
to any spin-dependent interaction. (One has to remember that a determination
of the fine structure constant can be done by many methods and some
of them do not involve any magnetic effects. In particular, one can
find $\alpha$ from $R_\infty$ and a certain $h/M$ value
\cite{chu,rb} as discussed in \cite{codata2006} and
\cite{previous}.) That allows to extend the constraint to larger
distances.

More detail on the data used for constraining the spin-dependent
long-range interaction from the $1s$ HFS can be found in
Appendix~\ref{s:a}.

\section{Extending the constraints to a larger-mass range}

\begin{figure}[thbp]
\begin{center}
\resizebox{0.95\columnwidth}{!}{\includegraphics{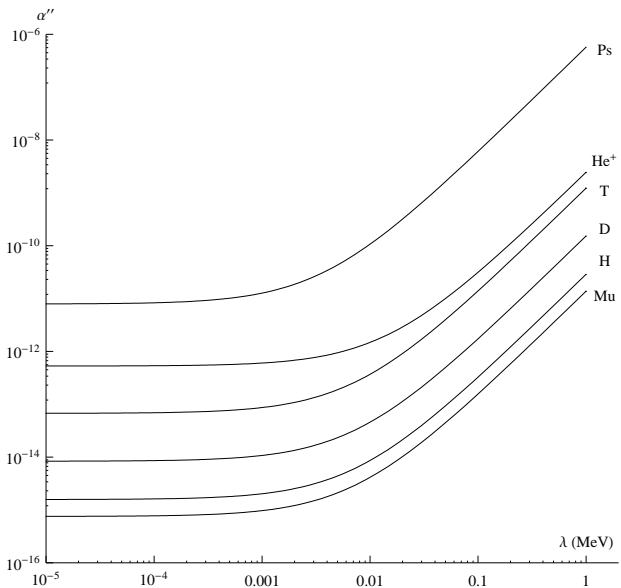}}
\end{center}
\caption{Constraint on a long-term spin-dependent interaction from
the HFS intervals of the ground state in light two-body atoms. The
lines present the upper bound on $|\alpha^{\prime\prime}|$ from data
on the $1s$ HFS interval in muonium, hydrogen, deuterium, tritium,
helium-3 ion, and positronium. A mass of an
intermediate particle $\lambda$ is the inverse Yukawa radius.
The confidence level is related to one standard deviation.}
\label{f:const}       
\end{figure}

A constraint from the value of the $1s$ HFS interval can be easily
extended to higher values of the mass of the intermediate boson,
$\lambda$. A direct calculation of the contribution of the Yukawa
spin-dependent term in (\ref{ssl}) leads to
\begin{equation}
\delta E_{\rm hfs}=-2\frac{\alpha^{\prime\prime}}{\alpha^2}
Z^2\,\frac{m_r}{m_e}\,R_\infty\times{\cal F}_1(\lambda/Z\alpha m_r)
\;,
\end{equation}
and so
\begin{equation}
\alpha^{\prime\prime}(\lambda)=\frac{\alpha^{\prime\prime}_0}{{\cal
F}_1(\lambda/Z\alpha m_r)}\;.
\end{equation}
Here $\alpha^{\prime\prime}_0$ is a related constraint for
$\lambda\ll Z\alpha m_e =3.5 Z\;$keV, listed in
Table~\ref{T:const1s}, and the profile function is of the form
\[
{\cal F}_1(x)=\left(\frac{2}{2+x}\right)^2\;.
\]
The constraint for the extended $\lambda$ range is presented in
Fig.~\ref{f:const} \cite{prl}. Here, the nuclear charge $Z$ is unity
for all atoms, but helium-3 ion ($Z_h=2$) and the reduced mass $m_r$
is equal to the electron mass $m_e$ for all atoms, but positronium
($m_r({\rm Ps})=m_e/2$). Because of that, the mass dependence of
positronium and helium constraints is somewhat different from
results derived from muonium, hydrogen, deuterium and tritium.

\section{Summary}

The final constraints from the $1s$ HFS in six different two-body
atoms are summarized in Fig.~\ref{f:const} \cite{prl}. For the
atomic systems, where the nuclear-effects are not well estimated, we
apply rather a rough estimation which, in principle, can be
substantially improved. The strength of constraint for
$\alpha^{\prime\prime}_0$ from muonium and positronium is determined
by experimental accuracy. For muonium that is the experimental
accuracy of determination of the muon magnetic moment in appropriate
units. For positronium the dominant uncertainty is the one for
measurements of the $1s$ HFS interval, while the theoretical
uncertainty is smaller, but still comparable with the experimental
one.

For hydrogen \cite{exph}, deuterium \cite{expd}, tritium \cite{expt}
and helium-3 ion \cite{exphe} the experimental data are
substantially more accurate, and the strength is determined by
an uncertainty in understanding the nuclear effects, which limits the
theoretical accuracy (see, e.g., \cite{my_rep}).

One can note that the  spin-dependent constraint obtained
here is stronger comparing with our spin-independent constraints
\cite{previous}, which, in fact, deal with substantially more
accurate data. A reason for that is that the interaction (\ref{ss})
modifies the Coulomb interaction and thus an enhancement factor
(i.e., a factor of $2\times10^8$ in (\ref{mucor})) comparing with
the magnetic interaction appeared.

It is also notable that the constrained boson is responsible for a
kind of interaction, which is somewhat similar to the weak
interaction by $Z$ boson exchange. While the weak contribution to
the $1s$ HFS in light hydrogen-like atoms is below the accuracy of
comparison of theory and experiment (see, e.g.
\cite{my_rep,EGS,codata2006}), our constraint is quite strong.

Concerning the weak interaction in atomic physics, we have to remind
that the weak interaction is weaker than the electromagnetic one not
because of a weak coupling constant, but because of the heaviness of
its intermediate particle. Introducing a new particle, the
correction would increase with a lighter mass and decrease with a
weaker coupling constant. From the point of view of the final
result, a correction with near-zero mass and an ultraweak
interaction can be compatible with a conventional weak interaction.
Indeed, that is possible only at the low momentum transfer. In
similar matter various atomic weak-interaction experiments can also
constrain certain long-range interaction.

Concluding, here, we constrain not just a pseudovector particle, but
a particle with spin one and axial coupling. That covers not only
pseudo vectors, but also a light particle, which similarly to $Z$
boson, does not have a fixed parity. The weak interaction
experiments in atomic physics are unlikely useful to constrain light
pseudovectors, however, it seems that constraining the pseudo vector
we also constraint corrections to the weak interaction at atomic
distances.

\section*{Acknowledgements}

This work was supported in part by RFBR (grants \#\# 08-02-91969 \&
08-02-13516) and DFG (grant GZ 436 RUS 113/769/0-3). The author is
grateful to Andrej Afanasev, Dmitry Toporkov, Eugene Korzinin, Simon
Eidelman, Maxim Pospelov, and Oleg Sushkov for useful and
stimulating discussions.

\appendix

\section{Summary on experimental and theoretical data on the $1s$ HFS interval in light
two-body atoms\label{s:a}}

Here, we collect reference data applied in the evaluation. The
experimental results on the $1s$ HFS interval are collected in
Table~\ref{t:exp1}. They represent physics at atomic scale.

\begin{table}[phtb]
\begin{tabular}{clc}
\hline
 ~~~~~~~~Atom~~~~~~~~ & ~~~~$E_{\rm HFS} ({\rm exp})$~~~~ & ~~~~Refs.~~~~\\
 & ~~~~~~~~[kHz] &  \\
 \hline
Muonium & 4\,463\,302.78(5) &  \cite{mu1shfs}\\
Hydrogen& 1\,420\,405.751\,768(1)&\cite{exph}\\
Deuterium & ~~327\,384.352\,522(2) &\cite{expd}\\
Tritium & 1\,516\,701.470\,773(8) &\cite{expt}\\
$^3$He$^+$ ion & - 8\,665\,649.867(10)&\cite{exphe}\\
Positronium & 203\,389\,100(740)  & \cite{ps}\\
\hline
\end{tabular}
\caption{The most accurate results for the $1s$ HFS interval in
light hydrogen-like atoms. A negative sign for the $^3$He$^+$ ion
reflects the fact that the nuclear magnetic moment is negative,
i.e., in contrast to other nuclei in the Table, its direction is
antiparallel to the nuclear spin.\label{t:exp1}}
\end{table}

To constrain a long-range interaction, one has to compare it with
physics on macroscopic distances, which provides a value of the
magnetic moment of the involved nuclei. We summarize in
Table~\ref{t:mu} results on $\mu_{\rm nucl}/\mu_B$, values of
magnetic moments of the nuclei of interest in units of the Bohr
magneton.

\begin{table}[phtb]
\begin{tabular}{clll}
\hline
Quantity & ~~~~~~Value & ~~~~~~~~Method& Refs.\\
&~~~~~~[$\times10^{-3}$]&&\\
 \hline
$\mu_\mu/\mu_{\rm B}$& $4.841\,970\,49(12)$ & BR Mu @ $B$ \& NMR p & \cite{mu1shfs}\\
$\mu_p/\mu_{\rm B}$&$1.521 \,032\, 209(12) $ & BR H @ $B$ & \cite{muep}\\
$\mu_d/\mu_{\rm B}$&$1.041 \,875 \,63(25) $ & BR D @ $B$& \cite{mued}\\
&&and BR H \& NMR HD&\cite{nmr}\\
$\mu_t/\mu_{\rm B}$&$1.622 \,393 \,657(21) $ & NMR HT & \cite{magtritium,nmr}\\
$\mu_h/\mu_{\rm B}$&$1.158\,741\,958(14)$ & NMR He \& H$_2$O  & \cite{maghelium,gammap}\\
 \hline
\end{tabular}
\caption{Determination of the nuclear magnetic moment in light
atoms. `BR @ $B$' is for study of the Breit-Rabi levels at presence
of magnetic field (in muonium, hydrogen and deuterium); `NMR' stands
for nuclear magnetic resonance of free protons (p), atoms of $^3$He
and molecules of HD, HT and H$_2$O. The references and description
of the method are given for the most crucial measurements only. `h'
stands for helion, the nucleus of the helium-3 atom. All results,
but result for the helion, the nucleus of $^3$He, are taken directly
from \cite{codata2006}. The helion result is obtained as explained
in the text. \label{t:mu}}
\end{table}

Determination of the nuclear magnetic moment for light atoms is
mostly done for bound nuclei. A number of measurements are done on
the Breit-Rabi levels in two-body atoms and the shielding correction
is discussed in \cite{muepth,codata2006}. The most complicated is
theory of the neutral helium-3 atom, a three-body system, where
theory with sufficient accuracy is presented in  \cite{shield}. Other
experiments are performed with the nuclear magnetic resonance
technique on diatomic molecules HD and HT. They are more
complicated for
calculations, however,  it is the ratio of nuclear
magnetic moments (deuterium-to-proton and tritium-to-proton) that
is measured and
this kind of isotopic calculations has relatively high accuracy (see
\cite{nmr} for detail).

All the references and description of the method are given in the
table for the most crucial measurements for each involved nucleus.
Other involved measurements were on the anomalous magnetic moments
of electron and muon and on the proton-to-electron mass ratio. All
three values are known with accuracy much better than required (see
\cite{codata2006} for detail). All results, but a result for the
helion, the nucleus of $^3$He, are taken from the CODATA tables of
recommended values \cite{codata2006}. To obtain the free helion
value we used the related shielded value of
$1.158\,671,471(14)\times10^{-3}$ from \cite{codata2006} and the
shielding factor ($\sigma=59.967\,43(10)\times10^{-6}$) recently
calculated in \cite{shield}.

\begin{table}[phtb]
\begin{tabular}{crl}
\hline
~~~~~~~ Atom~~~~~~~ & Fractional~ & ~~~~~~~Dominant source\\
 & uncertainty &  ~~~~~~~~~of uncertainty \\
 \hline
Muonium & 0.12 ppm~~& ~~~determination of $\mu_\mu/\mu_p$\\
Hydrogen& 1 ppm~~ & ~~~nuclear effects\\
Deuterium & 35 ppm~~ & ~~~nuclear effects\\
Tritium & 40 ppm~~ & ~~~nuclear effects\\
$^3$He$^+$ ion & 200 ppm~~ & ~~~nuclear effects\\
Positronium & 4.4 ppm~~ & ~~~experiment \& theory\\ \hline
\end{tabular}
\caption{Uncertainty of comparison of theory and experiment for the
$1s$ HFS interval in light hydrogen-like atoms. \label{t:unc1}}
\end{table}

To conclude a short overview of involved values and accuracies we
summarize in Table~\ref{t:unc1} the uncertainty of comparison of the
experiment and theory of the $1s$ HFS interval in light two-body
atoms. While the accuracy for muonium and positronium is well
understood and was numerously discussed in literature (see, e.g.,
\cite{my_rep}), the uncertainty for conventional atoms in the table
is rather a rough estimation accepted in this paper.


\begin{thebibliography}{00.}

\frenchspacing

\bibitem{mu1s2s}
V. Meyer, S. N. Bagayev, P. E. G. Baird, P. Bakule, M. G. Boshier,
A. Breitr\"uck, S. L. Cornish, S. Dychkov, G. H. Eaton, A.
Grossmann, D. H\"ubl, V. W. Hughes, K. Jungmann, I. C. Lane, Yi-Wei
Liu, D. Lucas, Y. Matyugin, J. Merkel, G. zu Putlitz, I. Reinhard,
P. G. H. Sandars, R. Santra, P. V. Schmidt, C. A. Scott, W. T.
Toner, M. Towrie, K. Tr\"ager, L. Willmann, and V. Yakhontov, Phys.
Rev. Lett. {\bf 84}, 1136 (2000).

\bibitem{mu1shfs} W. Liu, M. G. Boshier, S. Dhawan, O. van Dyck, P. Egan, X. Fei, M.
G. Perdekamp, V. W. Hughes, M. Janousch, K. Jungmann, D. Kawall, F.
G. Mariam, C. Pillai, R. Prigl, G. zu Putlitz, I. Reinhard, W.
Schwarz, P. A. Thompson, and K. A. Woodle, Phys. Rev. Lett. {\bf
82}, 711 (1999).

\bibitem{EGS} M.I. Eides, H. Grotch and V.A. Shelyuto, Phys. Rep. {\bf 342}, 63
(2001);\\ M.I. Eides, H. Grotch and V.A. Shelyuto, {\em Theory of
Light Hydrogenic Bound States\/}, Springer Tracts Mod. Phys. {\bf
222} (Springer, Berlin, Heidelberg, 2007).

\bibitem{my_rep}
S. G.~Karshenboim, Phys. Rep.  {\bf 422}, 1 (2005).

\bibitem{codata2006}
P. J. Mohr, B. N. Taylor, and D. B. Newell, Rev. Mod. Phys. {\bf
80}, 633 (2008).

\bibitem{weak} M.~I.~Eides, Phys. Rev.  A{\bf 53}, 2953 (1996).

\bibitem{rho}
J. R. Sapirstein, E. A. Terray, and D. R. Yennie, Phys. Rev. D{\bf
29}, 2290 (1984);\\
R.N.~Faustov, A.~Karimkhodzhaev and A.P.~Martynenko, Phys. Rev.
A{\bf 59}, 2498 (1999);\\
A. Czarnecki, S. I. Eidelman and S. G. Karshenboim, Phys. Rev. D{\bf 65} (2002);\\
S.I. Eidelman, S.G. Karshenboim and V.A. Shelyuto, Can. J. Phys.
{\bf 80}, 1297 (2002).

\bibitem{lbl} S.G. Karshenboim,  V.A. Shelyuto and A.I. Vainshtein,  Phys. Rev.
D{\bf 78}, 065036 (2008).

\bibitem{muon} K. Melnikov and A. Vainshtein, {\em Theory of the Muon Anomalous Magnetic Moment\/}.
Springer Tracts Mod. Phys. {\bf 216} (Springer, Berlin, Heidelberg,
2006);\\
 F. Jegerlehner, {\em The Anomalous Magnetic Moment of the
Muon\/}. Springer Tracts Mod. Phys. {\bf 222} (Springer, Berlin,
Heidelberg, 2008);\\
F. Jegerlehner and A. Nyffeler, Phys. Rep. {\bf 477}, 1 (2009).

\bibitem{holdom} B. Holdom, Phys. Lett. {\bf 166}B, 196 (1986).

\bibitem{pospelov} M. Pospelov, Phys. Rev. D{\bf 80}, 095002 (2009).



\bibitem{prl} S.G. Karshenboim, Phys. Rev. Lett. (2010) to be published; eprint arXiv:1005.4859.

\bibitem{previous} S.G. Karshenboim, eprint arXiv:1005.4872.

\bibitem{ps} M. W. Ritter, P. O. Egan, V. W. Hughes and K. A. Woodle, Phys. Rev.
A{\bf 30}, 1331 (1984);\\
A. P. Mills, Jr., and G. H. Bearman, Phys. Rev. Lett. {\bf 34}, 246
(1975);\\ A. P. Mills, Jr., Phys. Rev. A{\bf 27}, 262 (1983).

\bibitem{exph} H. Hellwig, R.F.C. Vessot, M. W. Levine, P. W. Zitzewitz, D. W.
Allan, and D. J. Glaze, IEEE Trans. IM-{\bf 19}, 200 (1970);\\
P.~W.~Zitzewitz, E. E. Uzgiris, and N. F. Ramsey, Rev. Sci. Instr.
{\bf 41},  81 (1970);\\
L. Essen, R. W. Donaldson, E.G. Hope and M. J. Bangham, Metrologia
{\bf 9}, 128 (1973);\\
D. Morris, Metrologia {\bf 7}, 162 (1971);\\
V. S. Reinhard and J. Lavanceau, in {\em Proceedings of the 28th
Annual Symposium on Frequency Control\/} (Fort Mammouth, N. J.,
1974), p. 379;\\
P. Petit, M. Desaintfuscien and C. Audoin, Metrologia {\bf 16}, 7
(1980);\\
J. Vanier and R. Larouche, Metrologia {\bf 14}, 31 (1976);\\
Y. M. Cheng, Y. L. Hua, C. B. Chen, J. H. Gao and W. Shen, IEEE
Trans. IM-{\bf 29}, 316 (1980);\\
S.~G.~Karshenboim, Can. J. Phys. {\bf 78}, 639 (2000).

\bibitem{expd} D. J. Wineland and N. F. Ramsey, Phys. Rev. {\bf 5}, 821 (1972).

\bibitem{expt} B. S. Mathur, S. B. Crampton, D. Kleppner and N. F. Ramsey, Phys. Rev. {\bf 158}, 14 (1967).

\bibitem{exphe} H. A. Schluessler, E. N. Forton and H. G. Dehmelt, Phys. Rev. {\bf 187}, 5 (1969).

\bibitem{muep} P.~F.~Winkler, D.~Kleppner, T.~Myint and F.~G.~Walther, Phys.
Rev. A \textbf{5}, 83 (1972); corrected according to a private
communication with D.~Kleppner as quoted in \cite{codata2006}.

\bibitem{mued} W. D. Phillips, D.~Kleppner, and F.~G.~Walther, private communication;
quoted according to \cite{codata2006}.

\bibitem{nmr} Yu. I. Neronov and S. G. Karshenboim, Phys. Lett. A{\bf 318}, 126
(2003).

\bibitem{magtritium} Yu. I. Neronov and A. E. Barzakh, Sov. Phys. JETP {\bf 45}, 871
(1977).


\bibitem{maghelium} J. L. Flowers, B. W. Petley, and M. G. Richards,  Metrologia {\bf
30}, 75 (1993).

\bibitem{aexp} D. Hanneke, S. Fogwell, and G. Gabrielse,
Phys. Rev. Lett. {\bf 100}, 120801 (2008).


\bibitem{chu}
A. Wicht, J. M. Hensley, E. Sarajlic, and S. Chu, Phys. Scr. T{\bf
102}, 82 (2002);\\
A. Wicht, E. Sarajlic, J. M. Hensley, and S. Chu, Phys. Rev. A{\bf
72}, 023602 (2005).

\bibitem{rb} M. Cadoret, E. de Mirandes, P. Clad\'e, S.
Guellati-Kh\'elifa, C. Schwob, F. Nez, L. Julien, and F. Biraben,
Phys. Rev. Lett. {\bf 101}, 230801 (2008).

\bibitem{gammap} W. D. Phillips, W. E. Cooke, and D. Kleppner,
Metrologia {\bf 13}, 179 (1977);\\
B. W. Petley and R. W. Donaldson, Metrologia {\bf 20}, 81 (1984).

\bibitem{muepth} S. G. Karshenboim and V. G. Ivanov, Phys. Lett. B {\bf 566}, 27 (2003).

\bibitem{shield} A. Rudzinski, M. Puchalski, and K. Pachucki, J. Chem.
Phys. {\bf 130}, 244102 (2009).

\end{thebibliography}
\end{document}